\documentclass{IEEEtran}
\usepackage{mathtools}
\usepackage{amssymb,amsthm,amsfonts,float,color}
\usepackage{nicefrac}
\usepackage{subfigure}
\usepackage{cite}
\usepackage{setspace}
\usepackage{times}
\usepackage{epsfig}
\usepackage{verbatim}
\usepackage{wrapfig}
\usepackage{amsmath}
\usepackage{enumitem}
\usepackage[normalem]{ulem}
\setlist{noitemsep}

\newtheorem{theorem}{Theorem}[section]

\newtheorem{lemma}[theorem]{Lemma}
\newtheorem{proposition}[theorem]{Proposition}

\newcounter{construction}[section]
\newenvironment{construction}[1][]{\refstepcounter{construction}\par\vspace*{.0in}
   \noindent {{\sc Construction~\theconstruction}. #1} \rmfamily}{\vspace*{.0in}}

\newcommand{\Mod}[1]{\,\textup{mod}\,#1}
\newcommand\remove[1]{}
\newcommand{\nc}{\newcommand}

\allowdisplaybreaks
\nc\bfa{{\boldsymbol a}}\nc\bfA{{\boldsymbol A}}\nc\cA{{\mathcal A}}\nc\sA{{\mathscr A}}
\nc\bfb{{\boldsymbol b}}\nc\bfB{{\boldsymbol B}}\nc\cB{{\mathcal B}}\nc\sB{{\mathscr B}}
\nc\bfc{{\boldsymbol c}}\nc\bfC{{\boldsymbol C}}\nc\cC{{\mathcal C}}\nc\sC{{\mathscr C}}
\nc\bfd{{\boldsymbol d}}\nc\bfD{{\boldsymbol D}}\nc\cD{{\mathcal D}}
\nc\bfe{{\boldsymbol e}}\nc\bfE{{\boldsymbol E}}\nc\cE{{\mathcal E}}
\nc\bff{{\boldsymbol f}}\nc\bfF{{\boldsymbol F}}\nc\cF{{\mathcal F}}\nc\sF{{\mathscr F}}
\nc\bfg{{\boldsymbol g}}\nc\bfG{{\boldsymbol G}}\nc\cG{{\mathcal G}}
\nc\bfh{{\boldsymbol h}}\nc\bfH{{\boldsymbol H}}\nc\cH{{\mathcal H}}
\nc\bfi{{\boldsymbol i}}\nc\bfI{{\boldsymbol I}}\nc\cI{{\mathcal I}}\nc\sI{{\mathscr I}}
\nc\bfj{{\boldsymbol j}}\nc\bfJ{{\boldsymbol J}}\nc\cJ{{\mathcal J}}
\nc\bfk{{\boldsymbol k}}\nc\bfK{{\boldsymbol K}}\nc\cK{{\mathcal K}}
\nc\bfl{{\boldsymbol l}}\nc\bfL{{\boldsymbol L}}\nc\cL{{\mathcal L}}
\nc\bfm{{\boldsymbol m}}\nc\bfM{{\boldsymbol M}}\nc\cM{{\mathcal M}}
\nc\bfn{{\boldsymbol n}}\nc\bfN{{\boldsymbol N}}\nc\cN{{\mathcal N}}
\nc\bfo{{\boldsymbol o}}\nc\bfO{{\boldsymbol O}}\nc\cO{{\mathcal O}}
\nc\bfp{{\boldsymbol p}}\nc\bfP{{\boldsymbol P}}\nc\cP{{\mathcal P}}\nc\eP{{\EuScriptP}}\nc\fP{{\mathfrak P}}
\nc\bfq{{\boldsymbol q}}\nc\bfQ{{\boldsymbol Q}}\nc\cQ{{\mathcal Q}}
\nc\bfr{{\boldsymbol r}}\nc\bfR{{\boldsymbol R}}\nc\cR{{\mathcal R}}\nc\sR{{\mathscr R}}
\nc\bfs{{\boldsymbol s}}\nc\bfS{{\boldsymbol S}}\nc\cS{{\mathcal S}}
\nc\bft{{\boldsymbol t}}\nc\bfT{{\boldsymbol T}}\nc\cT{{\mathcal T}}
\nc\bfu{{\boldsymbol u}}\nc\bfU{{\boldsymbol U}}\nc\cU{{\mathcal U}}
\nc\bfv{{\boldsymbol v}}\nc\bfV{{\boldsymbol V}}\nc\cV{{\mathcal V}}\nc\sV{{\mathscr V}}
\nc\bfw{{\boldsymbol w}}\nc\bfW{{\boldsymbol W}}\nc\cW{{\mathcal W}}\nc\sW{{\mathscr W}}
\nc\bfx{{\boldsymbol x}}\nc\bfX{{\boldsymbol X}}\nc\cX{{\mathcal X}}
\nc\bfy{{\boldsymbol y}}\nc\bfY{{\boldsymbol Y}}\nc\cY{{\mathcal Y}}
\nc\bfz{{\boldsymbol z}}\nc\bfZ{{\boldsymbol Z}}\nc\cZ{{\mathcal Z}}

\begin{document}

\title{Optimal LRC codes for all lenghts $n\le q$}

\author{\IEEEauthorblockN{Oleg Kolosov} \hspace*{.5in}
\and \IEEEauthorblockN{Alexander Barg} \hspace*{.5in}
\and \IEEEauthorblockN{Itzhak Tamo} \hspace*{.5in}
\and \IEEEauthorblockN{Gala Yadgar}
}

\maketitle

{\renewcommand{\thefootnote}{}\footnotetext{

\vspace{-.2in}
 
\noindent\rule{1.5in}{.4pt}

{Oleg Kolosov and Gala Yadgar are with Computer Science Department, Technion - Israel Institute of Technology, Haifa, Israel, emails olekol@gmail.com and gala@cs.technion.ac.il. 

Alexander Barg is with ISR/Department of ECE, University of Maryland, College Park, MD 20817, USA and with IITP, Moscow, Russia, email abarg@umd.edu. Research supported by NSF grants CCF1618603 and CCF1422955.

Itzhak Tamo is with School of Electrical Engineering, Tel Aviv University, Tel Aviv, Israel, email zactamo@gmail.com. Research supported by ISF grant no.~1030/15 and NSF-BSF grant no.~2015814.}
\vspace{-.1in}
}}

%
%

\renewcommand{\thefootnote}{\arabic{footnote}}
\setcounter{footnote}{0}

\begin{abstract}
A family of distance-optimal LRC codes from certain subcodes of $q$-ary Reed-Solomon codes, proposed by I.~Tamo and A.~Barg in 2014, assumes that the code length $n$ is a multiple of $r+1.$ By shortening codes from this family, we show that it is possible to lift this assumption, still obtaining distance-optimal codes. 
\end{abstract}


\section{Introduction}
Let $\cC$ be a $q$-ary code of length $n$ and cardinality $q^k$. We say that $\cC$ has locality $r$ if for every $i=1,\dots,n$
there exists a subset $I_i\subset\{1,\dots,n\}\backslash\{i\}, |I_i|=r$ such that for every codeword 
$\bfc=(c_1,\dots,c_n)$ and every $i=1,\dots,n$ the coordinate $c_i$ is a function
of the coordinates $\{c_i,i\in I_i\}.$ We call $\cC$ an $(n,k,r)$ LRC code, and call the subsets $A_i:=I_i\cup\{i\}$ {\em repair groups}.

Codes with the locality property were introduced in \cite{OnTheLocality}, which also proved the following upper bound on the minimum
distance of an $(n,k,r)$ LRC code:
  \begin{equation}\label{eq:sb}
  d_{\text{min}}(\cC)\le n-k-\Big\lceil \frac kr\Big\rceil+2.
  \end{equation}
We call an LRC code {\em optimal} if its distance is the largest possible given the other parameters. 
Several constructions of optimal LRC codes were proposed in the literature, among them   
\cite{prakash2012optimal,sil13,TPG13,OptimalLRC,LiMaXing17b,MaGe17,LiuMesnagerChen18}. In particular, \cite{OptimalLRC}
suggested a family of $q$-ary $(n,k,r)$ LRC codes for any $n\le q$ such that $(r+1)|n$ that are optimal with respect to the bound
\eqref{eq:sb}.
As shown recently in \cite{JinMaXing17}, in some cases it is possible to extend this construction to the case $n\le q+1$ (still assuming the divisibility).
  
The codes in \cite{OptimalLRC} are constructed as certain subcodes of Reed-Solomon (RS) codes. Namely, for a given $n$ 
the code is constructed as a subcode of the RS code of length $n$ and dimension $k+\lceil\frac kr\rceil-1.$
While the ``parent'' RS code is obtained by evaluating all the polynomials of degree $\le k+\lceil\frac kr\rceil-2,$ the LRC codes in \cite{OptimalLRC} are isolated by evaluating the subset of polynomials of the form 
   $$
   f_a(x)=\sum_{i=0}^{r-1}\sum_{j=0}^{\lceil\frac kr\rceil-1}a_{ij}g(x)^j x^i,
   $$
 where $\deg(f_a)\le k+\lceil\frac kr\rceil-2$ and where $g(x)$ is a polynomial constant on each of the repair groups $A_i.$
 
 As pointed out in \cite{OptimalLRC}, it is possible to lift the condition $(r+1)|n$, obtaining LRC codes whose distance is
 at most one less than the right-hand side of \eqref{eq:sb}. At the same time, \cite{OptimalLRC} did not give a concrete
 construction of such codes, and did not resolve the question of optimality. In this note we point out a way to 
 lift the divisibility assumption, constructing optimal LRC codes for almost all parameters. 
 
 Our results can be summarized as follows.
 \begin{theorem} \label{thm:main}
{Suppose that the following assumptions on the parameters are satisfied:
 
 (1) Let $s:=n\Mod (r+1)$ and suppose that $s\ne 1;$
 
(2)  Let 
     $$
    m=\Big\lceil\frac  n{r+1}\Big\rceil.
     $$
We assume that $\bar n:=m(r+1)\le q;$ 

 Then there exists an explicitly constructible $(n,k,r)$ LRC code $\cC$ whose distance is the largest possible for its parameters $n,k,$ and $r$.}
 \end{theorem}

{\em Remark:}
After this note was completed, we became aware that most of its results are implied by an earlier work by A. Zeh and 
E. Yaakobi \cite{ZY16}. Specifically, we prove a bound on the distance of LRC codes of
length $n$ given in Theorem \ref{thm:sb1}, which is sometimes stronger than the bound \eqref{eq:sb}. We also construct a family
of LRC codes obtained as shortenings of the codes in \cite{OptimalLRC} and 
use the bounds \eqref{eq:sb}, \eqref{eq:sb1} to show that they have the largest possible minimum
distance for their parameters. It turns out that our strengthened bound is a particular case of \cite[Thm.6]{ZY16},
and that the fact that shortening optimal LRC codes preserves optimality is shown in \cite[Thm.13]{ZY16}. 
This implies that the codes in \cite{OptimalLRC} can be shortened without sacrificing the optimality property.

In this note we give an explicit algebraic construction of the shortened codes from \cite{OptimalLRC}, which is not directly implied by \cite{ZY16}. We believe that the construction of codes presents some interest.
 We also give an independent, self-contained proof of the needed particular case of the bound on their distance. 
  
\section{The construction}
Let $F=\mathbb{F}_q,$ let $n$ be the code length and let $r$ be the target locality parameter. 
As stated above, we assume that $s\ne 1$ 
(the case $s=0$ accounts for the original construction in \cite{OptimalLRC} and is included below).
Let $t:=r+1-s.$

Let $\bar A\subset F$ be a subset of size $\bar n.$ Suppose that $\bar A$ is partitioned into disjoint
subsets of size $r+1$:
  \begin{equation}\label{eq:partition}
   \bar A=\bigcup_{i=1}^{m}{A_i}.
 \end{equation}
The set $A$ of $n$ coordinates of the code $\cC$ is formed of arbitrary $m-1$ blocks in this partition, say
$A_1,\dots,A_{m-1},$ and an arbitrary subset of the block $A_{m}$ of size $s$ (our construction includes the case of $(r+1)|n$ in \cite{OptimalLRC}, in which case this subset is empty). Denote by $B$ the subset
of $A_{m}$ that is not included in $A$, so that 
  $$
  A=\bigcup_{i=1}^{m-1}{A_i}\cup (A_{m}\backslash B).
  $$
Let $g(x)\in \mathbb{F}[x]$ be a polynomial of degree $r+1$ that is constant on each of the blocks of the partition \eqref{eq:partition} and let $\gamma$ be the value of $g(x)$ on the points in the set $A_{m}$. Without loss of generality we will assume that $\gamma=0$ (if not, we can take the polynomial $g(x)-\gamma$ as the new polynomial $g(x)$). A way to construct such
polynomials relies on the structure of subgroups of $F$ and was presented in \cite{OptimalLRC} (see also \cite{LiuMesnagerChen18}).

The codewords of $\cC$ are formed as evaluations of specially constructed polynomials $f(x)$ on the set of points $A.$
To define the polynomials, let $k':=k+t$ and define the quantity
   $$
   S_{k'\!,r}(i)=
\begin{cases}
\Big\lfloor \frac{k'}{r}\Big\rfloor & i<k' \Mod r \\[.05in]
\Big\lfloor \frac{k'}{r}\Big\rfloor-1 & i\geq k' \Mod r.
\end{cases}, i=0,\dots,r-1.
  $$
 Next let $a\in F^k$ be a data vector. Write $a$ as a concatenation of two vectors:
   \begin{equation}\label{eq:ar}
  \begin{aligned}
   (a_{ij},i=0,&\dots,r-1,j=1,\dots,S_{k'\!,r}(i))\\
   &(b_m,m=0,\dots,s-2).
   \end{aligned}
   \end{equation}
The total number of entries in the vectors in \eqref{eq:ar} equals
  \begin{multline*}
  (k' \Mod r)\Big\lfloor \frac{k'}{r}\Big\rfloor + r-(k' \Mod r)\Big\lfloor \frac{k'}{r} - 1\Big\rfloor  \\ = \Big\lfloor \frac{k'}{r}\Big\rfloor r - r + (k' \Mod r) +r -t = k' - t = k,
  \end{multline*}
so \eqref{eq:ar} is a valid representation of the $k$-dimensional vector $a.$

Given $a$, let us construct the polynomial
  \begin{equation}\label{eq:fa}
  f_{a}(x)=\sum_{i=0}^{r-1}f_i(x)x^i+h_{B}(x)\sum_{m=0}^{r-t-1}b_mx^m,
  \end{equation}
where $h_{B}(x)=\prod_{\beta\in B}(x-\beta),\deg(h_B)=t$ is the annihilator polynomial of  $B$ and
   $$
   f_i(x):=\sum_{j=1}^{S_{k'\!\!,r}(i)}a_{ij}g(x)^j.
   $$ 

Let $\{\alpha_1,\alpha_2,\dots,\alpha_n\}$ be the set of elements of $F$ that corresponds to the indices in the set $A$.
Define the evaluation map
  \begin{equation}\label{eq:ev}
  a\stackrel{\text{ev}}{\mapsto} c_a:=(f_a(\alpha_i),i=1,\dots,n).
  \end{equation}
Varying $a\in F^k$, we obtain a linear $(n,k)$ code $\cC$. We summarize the construction as follows.

\vspace*{.1in}\begin{construction}\label{def:code} For given $n,k,r$ the LRC code $\cC$ of length $n$ and dimension $k$ 
is the image of the linear map $\text{ev}:F^k\to F^n$ defined in \eqref{eq:ev}.
\end{construction}

\vspace*{.1in} We note that the code $\cC$ forms a shortening of the code in \cite{OptimalLRC} by the coordinates in $B$, so overall $\cC$ is a shortened subcode of the RS code of length
$\bar n$.

\section{Properties of the code $\cC$}

\subsection{Locality}
Let us show that the code $\cC$ has locality $r$. Let $i$ be the erased coordinate. If $i\in A_j,j=1,\dots,m-1,$ then consider the restriction $(f_a)|_{A_j}$ of the polynomial $f_a$ to the set $A_j,|A_j|=r+1$. From \eqref{eq:fa}, 
$\deg(f_a)|_{A_j}=r-1,$ so on the set $A_j$ it can be interpolated from its $r$ values. Once the polynomial $(f_a(x))|_{A_j}$ is
found, we compute the value $c_i=(f_a)|_{A_j}(\alpha_i),$ completing the repair task.

Now suppose that $i\in A_{m}\backslash B.$ We note that the restricted polynomial
$(f_a)|_{A_m}$ of degree at most $r-1$ such that
$$ f_a(\alpha)=(f_a)|_{A_m}(\alpha)=0 \text{ for any $\alpha \in B$}.$$
Now note that $|A_{m}\backslash B|=s,$ and that $s-1$ out of these coordinates are known. Together with the zero values
at the points of $B$ this gives $s-1+t=r$ known values, implying that it is possible to find
 the restricted polynomial $(f_a)|_{A_{m}}.$ 
Once this polynomial is computed, evaluating it at the point $\alpha_i$ again gives back the value of the missing coordinate.

\subsection{Dimension and distance}
\begin{lemma}{\rm \cite[Thm.2.1]{OptimalLRC}}
Let $\cC$ be an $(n,k,r)$ LRC code, then
\begin{equation}
k\leq n- \Big\lceil \frac{n}{r+1} \Big\rceil.
\label{eq:stam}
\end{equation}
\end{lemma}

The results about the parameters of the code $\cC$ are summarized in the following proposition.

\begin{proposition}\label{lemma:degree}
 Let $\cC$ be an LRC code with locality $r$ given by Construction \ref{def:code}. Then $\dim(\cC)=k$ and
  \begin{align}\label{eq:dist}
  d\ge n-k-\Big\lceil\frac{k+t}{r}\Big\rceil+2.
  \end{align}
\end{proposition}
\begin{IEEEproof}
We begin with bounding the degree of the polynomials $f_a(x)$ in \eqref{eq:fa}.
Suppose that $ r\!\!\not|\,k' $, then the maximum degree is
  \begin{align}
  \deg(f_a)&= \Big\lfloor \frac{k'}{r}\Big\rfloor (r+1) + (k' \Mod r) -1\nonumber\\
  & = k' + \Big\lfloor \frac{k'}{r}\Big\rfloor -1 \\
&= k' + \Big\lceil \frac{k'}{r}\Big\rceil -2. \label{eq:deg1}
  \end{align}
Now consider the case $r|k',$ namely,
  \begin{align}
  \deg(f_a)&\le \Big(\Big\lfloor \frac{k'}{r}\Big\rfloor -1 \Big)(r+1) + (r-1) \nonumber
\\&= k' + \Big\lceil \frac{k'}{r}\Big\rceil -2.\label{eq:deg2}
  \end{align}
To prove that $\dim(\cC)=k$ it suffices to show that the image of a nonzero $a\in F^k$ under the map \eqref{eq:ev} is nonzero.
We will prove an even stronger fact, namely that $\text{wt}(c_a)\ge 2$ for any $a\ne 0.$  We know that $f_a(x)$ has $t$ of its zeros in $B$, so the number of zeros in the 
set $\cup_{i=1}^{m-1} A_i\cup(A_m\backslash B)$ is at most 
  $$\deg(f_a)-t\leq k'+\lceil k'/r\rceil -2-t.$$
Noting that $\lceil\frac n{r+1}\rceil=\frac{n+t}{r+1}$ and using \eqref{eq:stam}, we obtain
   \begin{align*}
n-k-\Big\lceil \frac{k'}{r}\Big\rceil&\geq \Big\lceil \frac{n}{r+1} \Big\rceil-\Big\lceil \frac{k+t}{r}\Big\rceil \\
&\geq \Big\lceil \frac{n}{r+1} \Big\rceil-\Big\lceil \frac{n-\lceil \frac{n}{r+1} \rceil+t}{r}\Big\rceil\\
&= \Big\lceil \frac{n}{r+1} \Big\rceil-\Big\lceil \frac{n- \frac{n+t}{r+1}+t}{r}\Big\rceil\nonumber\\
&= \Big\lceil \frac{n}{r+1} \Big\rceil-\Big\lceil \frac{n+t}{r+1}\Big\rceil=0.
  \end{align*}
  Thus the number of nonzero values of $f_a(x)$ within the support of the codeword is at least two\footnote{The distance of $\cC$ is in fact greater than 2, as is shown in the second part of this lemma. The reason that we obtain 2 here is that we rely on a universal
  bound \eqref{eq:stam} for the rate of the code $\cC$ which is valid for all parameters.}.
Hence the mapping \eqref{eq:ev} is injective on $F^k,$ which proves that $\dim(\cC)=k.$
The weight of a nonzero vector satisfies $\text{wt}(c_a)\ge n-\deg(f_a),$ which together with \eqref{eq:deg1}-\eqref{eq:deg2}
 proves inequality \eqref{eq:dist} for the distance of $\cC.$
\end{IEEEproof}

\subsection{Optimality}
Finally let us prove that the constructed codes are distance-optimal.
The following upper bound on the distance of LRC codes tightens the bound \eqref{eq:sb} in some cases.
\begin{theorem}\label{thm:sb1}
Let $\cC$ be an $(n,k,r)$ LRC such that $s:=n\Mod (r+1)\neq 0,1$. Suppose that $\cC$ has 
$m:=\lfloor \frac{n}{r+1}\rfloor$ disjoint repair groups $A_i$ such that $|A_i|=r+1,i=1,\dots,m-1$ and $|A_m|=s.$

If either $r|k,$ or $r \!\!\not|\, k$ and $k\Mod r\geq s,$ then 
   \begin{equation}\label{eq:sb1}
   d_{\text{min}}(\cC)\leq n-k-\Big\lceil \frac{k}{r} \Big\rceil +1.
   \end{equation}
\end{theorem}

\begin{IEEEproof}
The minimum distance of a $q$-ary $(n,k)$ code (with or without locality) equals
  $$
  d=n- \max_{I\subseteq [n]}\{|I|:\cC_I<q^k\}.
  $$
Let $A_i'\subset A_i$ be an arbitrary subset of size $|A_i|-1$ and let  
   $$
   I'=A_1'\cup\dots\cup A'_{\lfloor \frac{k-1}{r} \rfloor}\cup A_m'.
   $$
Note that 
    \begin{equation}
|I'|=r \Big\lfloor \frac{k-1}{r} \Big\rfloor +s-1.
\label{stam}
   \end{equation}
If $r|k$ then, since $s\leq r$, \eqref{stam} becomes 
   $$
   k-r+s-1\leq k-1
   $$ 
Similarly, if $r\!\!\not|\,k$ and $k\Mod r\geq s,$ then \eqref{stam} becomes
   $$
   k-1-((k\Mod r)-1)+s-1\leq k-1.
   $$
Thus in either case $|I'|\le k-1$. 

If $|I'|<k-1,$ let us add to it arbitrary $k-1-|I'|$ coordinates which are not in the set 
$$A_1\cup\dots\cup A_{\lfloor \frac{k-1}{r} \rfloor}\cup A_m,$$ again calling the resulting subset $I'.$
By construction, $|C_{I'}|\leq q^{k-1}.$ 

Now consider a larger subset of coordinates that is formed of the complete repair groups and the set $I'$,
  $$
  I=I'\cup A_1\cup\dots\cup A_{\lfloor \frac{k-1}{r} \rfloor}\cup A_m.
  $$
Because of the locality property, the coordinates in $I$ depend on the coordinates in 
$I',$ and so $|\cC_I|=|\cC_{I'}|.$
 Finally, $|I|=|I'|+\lfloor \frac{k-1}{r} \rfloor +1=k+\lceil \frac{k}{r} \rceil -1$. 
Therefore the minimum distance is at most 
$d\leq n-|I|,$ giving \eqref{eq:sb1}.
\remove{comment: the proof is not fully complete, because we need to argue using the upper bound on $k$ that there exists enough coordinates that we can add to the set $I'$ in order to make it of size $k-1$. But I think we can sweep it under the rug.}
\end{IEEEproof}

\vspace*{.1in}
\begin{proposition}
The codes given by Construction \ref{def:code} have the largest possible value of the distance for their parameters.
\end{proposition}
\begin{IEEEproof}
The distance of the code $\cC$ is given in \eqref{eq:dist}. The difference between $n-k-\lceil\frac kr\rceil+2$ in \eqref{eq:sb} and the right-hand side
of \eqref{eq:dist} equals
  $$
  \Delta:=\Big\lceil\frac{k+t}{r}\Big\rceil-\Big\lceil\frac kr\Big\rceil\le 1
  $$
  (since $t\le r-1$).

If $\Delta=0$, then the code $\cC$ is optimal by \eqref{eq:sb}. 

Let us prove that $\cC$ is optimal even when $\Delta=1.$ Let us find the parameters for which this holds true.
Let $k=ru+v,$ where $0\le v\le r-1.$ If $v=0$ (i.e., $r | k$), then clearly $\Delta=1.$ Otherwise, suppose 
that $v\ge 1$ and compute
  \begin{align*}
  \Big\lceil\frac{k+t}{r}\Big\rceil-\Big\lceil\frac kr\Big\rceil
  =\Big\lceil u+1+\frac{v+1-s}{r}\Big\rceil-(u+1),
  \end{align*}
which equals $1$ if and only if $v+1-s>0,$ i.e., if and only if $v=k\Mod r\ge s.$

In summary, $\Delta=1$ if and only if either $r|k,$ or $r \!\!\!\not|\, k$ and $k\Mod r\geq s.$
However, in both these cases, according to \eqref{eq:sb1}, the maximum possible distance 
is one smaller than the bound \eqref{eq:sb}. This again establishes optimality of
the code $\cC.$
\end{IEEEproof}

{\em Remark:}
In conclusion we note that the code family in \cite{OptimalLRC} affords an easy extension to the case when each repair group is resilient to more than one erasure (i.e., it supports a code with distance $\rho>2$). The construction in \cite{OptimalLRC} assumes that
$(r+\rho-1)|n.$ Using the ideas in the previous section, specifically,  polynomials of the form \eqref{eq:fa}, it is easy to lift this 
assumption, obtaining codes that support local correction of multiple erasures for 
any length $n\le q$ such that $\lceil \frac n{r+\rho-1}\rceil(r+\rho-1)\le q.$

{\bibliographystyle{abbrv}
	\bibliography{ErasureCodes_bib}
}

\end{document}